\documentclass[fleqn,twoside]{article}
\usepackage{espcrc2,psfig,epsfig}

\title{Lattice Study of the O(3) Supersymmetric Sigma Model}

\author{Sofiane Ghadab\address{Department of Physics\\ 
        Syracuse University, Syracuse NY 13244}%
        \thanks{research supported in part by DOE grant DE-FG02-85ER40231}}
       
\begin{document}

\begin{abstract}
We present preliminary numerical results from a lattice study of the
two-dimensional O(3) non-linear sigma
model. In the continuum this model possesses $N=2$ supersymmetry. 
The lattice formulation we use retains an exact (twisted)
supersymmetry except for a soft breaking associated with a Wilson mass
term needed to remove the doubles. Our numerical results show that
the partition function is independent of coupling as predicted by supersymmetry.
\vspace{1pc}
\end{abstract}

\maketitle

\section{General 2D Sigma Model}
As was first noted by Polyakov \cite{polyakov}, there exists a 
deeply rooted analogy between four-dimensional Yang-Mills theories and 
two-dimensional sigma models. Indeed the latter can serve as perfect 
theoretical laboratories to test methods and approaches developed for solving 
the problems of these far richer and more complicated theories. This is
particularly true for the supersymmetric extensions of these
theories \cite{novikov}. It is clearly important to formulate such
theories on the lattice and a greal deal of progress has been
made in this area recently for models possessing extended supersymmetry
\cite{wz,top,sig,sugino}. These formulations all rely on the idea of
{\it twisting} -- a technique pioneered by Witten \cite{witten} in the context
of topological field theory. These twisted formulations possess a
scalar nilpotent supercharge $Q$, which with care, may often be preserved
on the lattice.
This symmetry can help prevent 
relevant susy breaking counterterms from appearing in the effective action and 
thus may avoid the fine tuning problems of generic discretizations
of supersymmetric models. As 
was discussed in Simon Catterall's talk in this conference 
the twisted action is typically $Q$ exact and in the case of
the 2D sigma model takes the form
\[
S=\alpha Q\int_\sigma \eta_{i\mu}\left(N^{i\mu}\left(\phi\right)-
\frac{1}{2}g^{ij}B_{j}^{\mu}\right)
\]
where the action of the twisted charge on the fields is given by
\begin{eqnarray}
Q\phi^i&=&\lambda^i \nonumber \\
Q\lambda^i&=&0 \nonumber \\
Q\eta_{i\mu}&=&\left(B_{i\mu}-
\eta_{j\mu}\Gamma^j_{ik}\lambda^k\right) \nonumber \\
Q B_{i\mu}&=&\left(B_{j\mu}\Gamma^j_{ik}\lambda^k-\frac{1}{2}\eta_{j\mu}
R^j_{ilk}\lambda^l\lambda^k\right)\nonumber
\end{eqnarray}
Here $\phi^i(\sigma)$ is a set of commuting fields which act as
coordinates on some $N$-dim target space with metric $g_{ij}(\phi)$. 
$\lambda^i(\sigma)$ are scalar grassmann fields, $\eta_{i\mu}(\sigma)$ are
vector grassman fields and $B_{i\mu}(\sigma)$ are another commuting vector fields. 
If we carry out the $Q$-variation and integrate out $B$ we find
\begin{eqnarray}
S&=&\alpha\int_\sigma
\left(\frac{1}{2}g_{ij}N^{i\mu}N^{j}_{\mu}-\eta_{i\mu}
\nabla_kN^{i\mu}\lambda^k\right.\\
&+&\left.
\frac{1}{4}R_{jlmk}\eta^{j\mu}\eta^{l}_{\mu}\lambda^m\lambda^k\right)
\end{eqnarray}
To generate a kinetic term for the $\phi$ fields it is necessary to
choose
\[
N^{i\mu}=\partial^\mu\phi^i
\]
Finally to make contact with usual sigma model we need to impose
{\it self-duality} conditions on the vector fields in the model. This
amounts to replacing $\eta$ by $P^{(+)}\eta$ where
\[
{P^{i\mu}_{j\nu}}^{\left(+\right)}=
\frac{1}{2}\left(\delta^i_j\delta^\mu_\nu+
J^i_j\epsilon^\mu_\nu\right)
\]
and $J$ is a covariantly constant matrix $\nabla_k J^i_j=0$. Manifolds
possessing such a tensor field are termed K\"{a}hler. It can be shown that the
resultant model can we written in complex coordinates as
\begin{eqnarray}
S&=&\alpha\int d^2\sigma\left(
2h^{+-}g_{I\overline{J}}\partial_+\phi^I
\partial_-\phi^{\overline{J}}\right.\nonumber\\
&-&\left.h^{+-}g_{I\overline{J}}\eta^I_+D_-\lambda^{\overline{J}}-
h^{+-}g_{\overline{I}J}\eta^{\overline{I}}_-D_+\lambda^J\right.\nonumber\\
&+&\left.
\frac{1}{2}h^{+-}R_{I\overline{I} J\overline{J}}
\eta^I_+\eta^{\overline{I}}_-\lambda^J\lambda^{\overline{J}}\right)\nonumber
\end{eqnarray}

\section{2D Sigma Model on the lattice}
It is trivial to see that the action is invariant under the twisted 
supersymmetry when I replace the continuum derivatives by symmetric 
finite differences, as the continuum Q-symmetry makes no reference to 
derivatives of the fields. However the kernel of the (free) lattice 
Dirac operator constructed this way contains both fermionic and bosonic
doubles which have no continuum interpretation. To remove these we need to
add a Wilson term to the lattice action. Of course it is not obvious that 
the addition of such a term is compatible with the topological symmetry 
in the case of a curved target space. However, in \cite{wilson2} it 
was shown that indeed it is possible to add potential terms to the twisted 
sigma models while maintaining the topological symmetry. The possible terms are
\[
\Delta S= c^2
V^IV_I+c^2\lambda^I\nabla_IV_{\overline{J}}\lambda^{\overline{J}}-
\frac{1}{4}h^{+-}\eta_+^I\nabla_IV_{\overline{J}}\eta_-^{\overline{J}}
\label{killing}
\]       
Here, $V^I$ is a holomorphic Killing vector and $c^2$ an
arbitrary parameter. A Wilson term would correspond to
the choice $V^I=im_W\phi^I$ where
\[
m_W=\frac{1}{2}\left(\Delta^+_z\Delta^-_{\overline{z}}+
                     \Delta^+_{\overline{z}}\Delta^-_z\right)
\]
Many K\"{a}hler manifolds 
possess such a holomorphic Killing vector (for example the
$CP^N$ theories). Actually $Q^2$ is no longer zero in such models
but yields a diffeomorphism along the Killing vector. For a reparametrization
invariant continuum action this will be an exact symmetry. However on
the lattice (where exact reparametrization invariance is broken) it will
yield a symmetry breaking term of $O(a^3)$. As discussed in
\cite{sig} such a term will not affect
the renormalization of the model and we expect that the model will
flow to a continuum theory with full $N=2$ SUSY in the continuum limit. 

\section{Lattice O(3) nonlinear sigma  model}

The
usual $O(3)$ supersymmetric sigma model is perhaps the simplest
example of a model with K\"{a}hler target space which may be realized
as a twisted model. 
In this case the metric, connection and
curvature are easily verified to be
\begin{eqnarray}
g_{u\overline{u}}&=&\frac{1}{2\rho^2}\\
\Gamma^u_{uu}&=&g^{\overline{u}u}\partial_u g_{\overline{u}u}=-\frac{2\overline{u}}{\rho}\\
R_{\overline{u}u\overline{u}u}&=&g_{\overline{u}u}\partial_{\overline{u}}\Gamma^u_{uu}=-\frac{1}{\rho^4}
\end{eqnarray} 
where $\rho=1+u\overline{u}$.
In this case the supersymmetric lattice action including Wilson
terms takes the form
\begin{eqnarray}
S&=&\alpha\sum_{x}\left[\frac{1}{\rho^2}\Delta^S_+u\Delta^S_-\overline{u}
+\frac{1}{\rho^2}(m_Wu)(m_W\overline{u})\right.\nonumber\\
&&-\frac{1}{2\rho^2}\eta D^S_-\overline{\lambda}-\frac{1}{2\rho^2}\overline{\eta} D^S_+\lambda
-\frac{1}{2\rho^4}\overline{\eta}\eta\overline{\lambda}\lambda\nonumber \\
&&+\frac{1}{2\rho^2}\lambda i[m_W-\frac{2\bar{u}}{\rho}(\Delta_+^Su)+c.c]\overline{\lambda}\nonumber \\ 
&&-\left. \frac{1}{8\rho^2}\eta
i[m_W-\frac{2\bar{u}}{\rho}(\Delta_+^Su)+c.c]\overline{\eta}
\right]
\end{eqnarray}
To proceed further it is convenient to introduce an auxiliary field $\sigma$
to remove the quartic fermion term. Explicitly we employ the
identity
\begin{equation}
\alpha^N\int D\sigma e^{-\alpha\left(\frac{1}{2}\sigma\overline{\sigma}+
\frac{\sigma}{2\rho^2}\overline{\eta}\lambda+
\frac{\overline{\sigma}}{2\rho^2}\eta\overline{\lambda}\right)}=
e^{\frac{\alpha}{2\rho^4}
\overline{\eta}\eta\overline{\lambda}\lambda}
\end{equation}
where $N$ is the number of lattice sites. Thus the partition function
of the lattice model can be cast in the
form
\begin{equation}
Z=\int DuD\sigma D\eta D\psi e^{-S\left(u,\sigma,\eta, \lambda\right)}
\end{equation}
where the action is now given by
\begin{eqnarray}
S&=&\alpha\sum_{x}\left[\frac{1}{\rho^2}\Delta^S_+u\Delta^S_-\overline{u}+
\frac{1}{\rho^2}(m_Wu)(m_W\overline{u}) \right. \nonumber \\
&&+\left.\frac{1}{2}\sigma\overline{\sigma}+
\overline{\Psi} M(u,\sigma) \Psi \right]
\label{o3act}
\end{eqnarray}
The anticommuting fields are assembled into Dirac spinors 
\[\begin{array}{ccc}
\Psi=\left(\begin{array}{c}\bar{\lambda}\\
\frac{1}{2i}\bar{\eta}\end{array}\right)&\;\;\;&
\overline{\Psi}=\left(\begin{array}{c}\lambda\\
\frac{1}{2i}\eta\end{array}\right)
\end{array}
\]
and the Dirac operator $M(u,\sigma)$ in the chiral basis is
\[M(u,\sigma)=\frac{i}{2\rho^2}\]
\[\left(\begin{array}{cc}
m_W-\frac{2\bar{u}}{\rho}m_Wu+c.c & D_+\\
-D_+^\dagger & m_W-\frac{2\bar{u}}{\rho}m_Wu+c.c
\end{array}\right)\]
where the covariant derivative is modified to include a coupling to
the auxiliary field $\sigma$
\begin{equation}
\hat{D}_+=\Delta^S_+ -\frac{2\overline{u}}{\rho}\left(\Delta^S_+
u\right)+\sigma
\label{cov}
\end{equation}
To simulate this model we have to reproduce the
fermion determinant arising after integrating out the
anticommuting fields. This leads to an effective action of the
form (we observe the determinant to be positive definite)
\begin{equation}
S=\alpha S_B(u,\sigma)-\frac{1}{2}{\rm Tr}
\ln{\left(\alpha^2M^\dagger(u,\sigma)M(u,\sigma)\right)}
\end{equation}
where $S_B(u,\sigma)$ denotes the local bosonic pieces of the action and
we have shown the dependence on coupling $\alpha$ explicitly.
We have employed a Langevin algorithm
with a stochastic fermion force estimator to
represent this determinant. In this
case physical quantities exhibit systematic errors 
of order the step size $O(\tau)$.

\begin{figure}
\epsfig{file=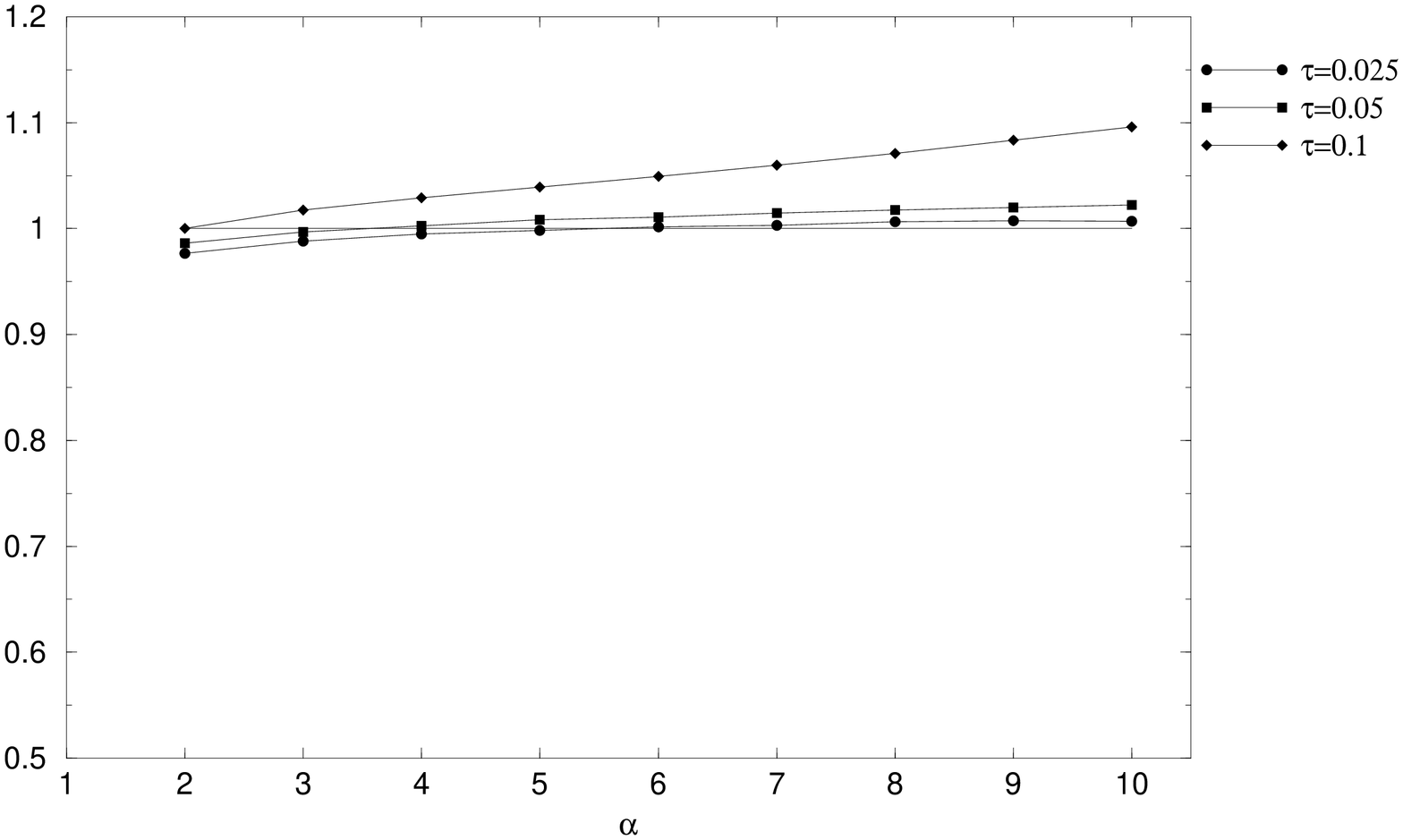,angle=-90,width=7.5cm}
\caption{$\alpha<S_B>/2V$}
\label{action}
\end{figure}

Using the invariance of the action under the twisted $Q$, one can easily 
show that the partition function should be {\it independent} of the coupling
constant $\alpha$. A stringent test of this $\alpha$ independence of $Z$ is gotten by
measuring the expectation value of $S_B$. It should be clear that
\begin{equation}
-\frac{\partial\ln Z}{\partial\alpha}=0=\left<S_B\right>-\frac{2V}{\alpha}
\end{equation}
where $V$ denotes the number of lattice sites. The results are given in fig.\ref{action} 
which shows data
for $\frac{\alpha}{2V}<S_B>$ from runs at a variety
of couplings $\alpha$ for a lattice of size 12x12 for different time steps $\tau$.   
This is in a good agreement with the theoretical prediction above. Notice that the 
small deviations from unity at small $\alpha$'s are due to small Q-symmetry breaking 
associated with the Wilson term, whereas those for
large
$\alpha$'s are due to the systematic errors ($O(\tau)$) in the Langevin algorithm.
Further numerical studies of Ward identities and other important quantities are 
underway. They should give us a better understanding of the theory and the possibility 
of breaking of the supersymmetry.

\end{document}